\documentclass{article}

\usepackage{arxiv}

\usepackage[utf8]{inputenc} % allow utf-8 input
\usepackage[T1]{fontenc}    % use 8-bit T1 fonts
\usepackage{hyperref}       % hyperlinks
\usepackage{url}            % simple URL typesetting
\usepackage{booktabs}       % professional-quality tables
\usepackage{amsfonts}       % blackboard math symbols
\usepackage{nicefrac}       % compact symbols for 1/2, etc.
\usepackage{microtype}      % microtypography
\usepackage{lipsum}
\usepackage{amsmath}
\usepackage{graphicx}
\usepackage[authoryear]{natbib}
\usepackage[super]{nth}     % n-th
\usepackage{setspace}   %Allows double spacing with the \doublespacing command

\usepackage{mathptmx}
\usepackage{multirow}

\title{Diversity and Density of Urban Functions in Station Areas}

\author{
Yusuke Kumakoshi\thanks{Corresponding author}\\
Research Center for Advanced Science and Technology\\
University of Tokyo\\
Tokyo, 153-8904, Japan\\
\texttt{ykuma@cd.t.u-tokyo.ac.jp}\\

\And
Hideki Koizumi\\
Department of Urban Engineering\\
University of Tokyo\\
Tokyo, 113-8656, Japan\\
\texttt{hide@cd.t.u-tokyo.ac.jp}\\

\And
Yuji Yoshimura\\
Research Center for Advanced Science and Technology\\
University of Tokyo\\
Tokyo, 153-8904, Japan\\
\texttt{yyyoshimura@cd.t.u-tokyo.ac.jp}\\
}

\begin{document}
%  \pagenumbering{gobble} % erase page numberings
\maketitle

\doublespacing 

\begin{abstract}
The diversity and density of urban functions have been known to affect urban vibrancy positively, but the relation between the two has not been empirically examined; if high density is associated with low diversity in an area, its vibrancy may not increase. To obtain a better understanding of the metabolism of cities and directions for urban planning interventions, this paper offers empirical evidence on the association between the diversity and density of urban functions in the Tokyo Metropolitan Area, using a robust density index that was determined via a Monte Carlo simulation. By conducting association analyses, it was found that highly dense station areas tended to display low diversity at multiple scales. Further investigation indicated that this negative correlation was owing to different spatial characteristics of functions and complementary functioning among highly accessible station areas. This paper argues for considering both diversity and density in urban planning to make station areas vibrant and resilient.
\end{abstract}

% keywords can be removed
\keywords{Diversity, Density, Station area, Urban function, Tokyo}

\section{Introduction}
Urban functions that meet citizens' needs are fundamental to urban vibrancy, which enables sustainable urban development (\cite{hall2013urban}), because they attract people (\cite{wu2018checkin}). In this regard, urban planners have implemented two principles in cities: diversity and density. Diversity of functions, or mixed land use, has been recognized as a promoter of activities (\cite{Calthorpe1994,jacobs1961death,register1987ecocity}), and researchers have investigated its positive effects on urban life (\cite{dovey2017functional,song2013comparing}), including transportation ridership (\cite{cervero1997travel,ewing2010travel,sung2011transit}); public health (\cite{stevenson2016land}); and economic growth or vibrancy (\cite{duranton2000diversity,glaeser1992growth,quigley1998urban}). Density of functions, in turn, is expected to have a positive impact on an area because it produces a centeredness of the city (\cite{hester2010design}), and, from an economic perspective, stimulates economic efficiency via Marshallian externality (\cite{duranton2004micro}).

Theoretically, the densification of facilities is caused by consumer preference for diversity of products or services (\cite{fujita1996economics}). If this is empirically true, the eventual clustering of the facilities would be diverse and dense simultaneously, thereby vitalizing the urban area in terms of both aspects. However, little is empirically known about the interaction between these elements—that is, whether they occur independently or concomitantly and, if the latter, whether they are mutually reinforcing or inhibiting. If high density of functions is not associated with high diversity, urban planners must design restrictions to control the degree of density and diversity, and develop strategies for vitalization.

In effect, the presence of the negative externality of high density, such as congestion or pollution, casts doubt on a linear relation between diversity and density of urban functions. Although \cite{talen2006neighborhood} focused on the sociodemographic aspect, she found that population density affects income diversity nonlinearly; that is, excessive density results in less diversity. From an economic perspective, a study of individual firms' strategies suggests that the diversity of industrial sectors would decrease as the degree of agglomeration intensifies (\cite{alcacer2014location}). Conversely, \cite{duranton2000diversity} argued that large cities tend to be more diversified than small cities in terms of industry at the city level, thus supporting the simultaneous intensification of diversity and density. However, no study has empirically investigated the relation between the two in the neighborhood scale. %The present study provides empirical evidence on the relation among urban functions to obtain insights into the mechanism behind such interplays, and accordingly devise management systems that benefit urban areas.

Furthermore, when examining these links, using a robust diversity index to sample size is crucial; if the index value increases as the number of facilities increases, diversity \textit{per se} cannot be separately measured from density. Although studies have traditionally used Shannon's entropy (\cite{shannon1948mathematical}) to measure land use (\cite{cervero1997travel,sung2011transit}) and point-of-interest-based mixed use (\cite{yue2017measurements}), biological and ecological scholars have criticized its sensitivity to sample size (\cite{magurran1988ecological,Morishita1996}). In regard to sensitivity, Simpson's index (\cite{simpson1949measurement}), which is less frequently used in urban settings (\cite{arribas2011multidimensional,lowry2014comparing}), has been recommended (\cite{mouillot1999comparison,song2013comparing}), but its robustness has been verified in hypothetical populations of biological species and not those of urban functions. Hence, its suitability for urban settings remains unclear.

To fill these gaps in the literature, the present study empirically analyzes the relation between the diversity and density of urban functions by focusing on railway station areas (that are at a walkable distances from stations) in the Tokyo Metropolitan Area, Japan. The authors tested the hypothesis that a higher diversity of urban functions is associated with a higher density of those functions, using two diversity indices; the appropriateness of the indices in urban settings was also examined using a Monte Carlo simulation. The results provide insights into the mechanisms guiding such interplays, and accordingly help develop management systems that benefit urban areas.

This paper makes two contributions. First, it clarified that the diversity and density of urban functions—both being sources of urban vibrancy—do not mutually reinforce at the neighborhood scale. This suggests that urban planners should control the two parameters simultaneously to render an area vibrant; this point has not been claimed in previous studies. Second, it validated two diversity indices to measure urban functions' mixture via a Monte Carlo simulation. This allows future studies to use the indices in the same context.
%after related studies are reviewed in Section \ref{section:2},
% The remainder of the paper is organized as follows: the research methodology is explained in Section \ref{section:2}; the results are then detailed in Section \ref{section:3} and  discussed in Section \ref{section:4}; and finally, Section \ref{section:5} concludes the paper.

%%%%%%%%%%%%%%%%%%%%%%%%%%%%%%%%%%%%%%%%%%
\section{Data}
\label{section:2}

\subsection{Study area}
Empirically testing a hypothesis regarding the urban functions in station areas requires a wide range of stations in terms of scale and characteristics for statistical reliability. From this viewpoint, the Tokyo Metropolitan Area, one of the world's largest metropolises (\cite{florczyk2019ghs}), is suitable for this study. During the \nth{20} century, the metropolitan area experienced rapid population growth, and railway operators developed railway networks to meet the commute demand. Most of the railway lines spread radially out from their termini at intersections with the Yamanote loop line, which connects urban cores such as Shibuya and Shinjuku. These railway lines comprise a huge, complex railway network together with the subway lines that mainly serve the inner area of the Yamanote line (\cite{sorensen2001subcentres}) and contain stations of varied sizes. Considering this history of the area's development, this study investigated the stations at the suburban area outside the Yamanote line, which are often regarded as the center of development (\cite{cervero1998transit,chorus2016developing}).

In accordance with the literature on transit-oriented development (TOD) (\cite{cervero1997travel}), this study focused on station areas, which are easily accessible from stations on foot. Uniquely sized buffers (400 m) were defined for each studied station. The size was set at a constant for each station in spite of their heterogeneity (i.e., the number of railway lines and the ridership) because the walkable distance can be assumed to be relatively stable, regardless of the place where people walk.

\subsection{Dataset and preprocessing}
\label{sec:cat_def}
\textbf{Facilities}: This paper defined a ``facility" as a public or private entity that performs one type of function. This definition is broader than that of the ``firm" in economics (\cite{frenken2007related,fujita1996economics}), in that one such firm may consist of more than one facility if it performs multiple types of functions. This paper's definition also includes non-business organizations, such as public schools and temples. Such a definition is appropriate for the research because it considers all station area functions, regardless of the providers' characteristics. % definition of facilities

The facilities data were derived from a telephone directory that was provided by Zenrin Co., Ltd. The list of facilities was georeferenced and contained information on types of activities in 2018. Entries with multiple phone numbers were placed under one facility if they contained identical names (e.g., an entity may possess multiple phone numbers). If there were different names and types under one phone number, they were regarded as separate facilities because they performed different functions (e.g., a temple and a hotel). The dataset covers a majority of the existing facilities, representing approximately 70\% of the total number of existing facilities in Shibuya ward, Tokyo, compared to the number of juridical persons registered in the Economic Census 2016. % More detail necessary? Slight difference between these definitions...

\textbf{Stations}: This study investigated 1,352 railway stations in Tokyo Metropolitan Area at the end of 2018. Data were retrieved from the National Land Information Platform of the Ministry of Land, Infrastructure, Transport, and Tourism (\url{https://nlftp.mlit.go.jp/ksj/gml/datalist/KsjTmplt-N02-v2_3.html}). In total, 96 lines (19 operators) were included in the analysis. Excluded stations were those located in the interior of the Yamanote loop line that had no transfer connections to the Yamanote line stations. Transport systems with an inferior traffic capacity to railways (e.g., monorails and the automated guideway transit) were also excluded from the analysis. % why this exclusion?
% central point of stations?

\textbf{Definition of urban function categories}: Facilities were categorized into six general categories and 26 sub-categories (Table S1 in the supplementary material), based on previous works on economic activities (\cite{porta2012street,yoshimura2020spatial}). These works focused more on the facilities' spatial dimensions, similar to the present study, and less on economic output (e.g., \cite{quigley1998urban}). The aforementioned works also showed a high interpretability of each general category and sub-category, without too excessively specifying particular facility types. This consistency with previous works on urban settings will enable scholars to adopt the categorical framework in other local settings, regardless of region.
%Biological diversity is based on established concepts of species, whereas the diversity of urban functions (or facilities) requires a clear definition of categories. 

%%%%%%%%%%%%%%%%%%%%%%%%%%%%%%%%%%%%%%%%%%
\section{Methodology}
\label{section:3}

\subsection{Measurement of diversity and density}
\textbf{Diversity}: Diversity of industry (e.g., Hirschmann–Herfindahl index), or land use mix, is typically measured with Simpson's diversity index (\cite{simpson1949measurement} or Shannon's entropy (\cite{shannon1948mathematical}). These measures are summarized to a single expression for diversity, called Hill number (\cite{hill1973diversity}; Eq \ref{eq:hill}), with different values for the parameter \textit{q}, which controls the weights of rare species.

\begin{equation}
    ^qD = (\sum_{i}^S p_i^q)^{\frac{1}{1-q}}
    \label{eq:hill}
\end{equation}

$^qD$ is the diversity of the degree of parameter $q$, and $p_i$ is the proportion of the number of individuals $i$ to the total number of individuals in the sample. Simpson's index corresponds to the case $q=2$, which weights rare species less and dominant species more than Shannon's entropy does ($q=1$). The authors selected these two indices in this study and not $q=0$, which corresponds to the number of species present in a sampled group. This exclusion is because its capacity of distinction between station areas is limited: it cannot distinguish two station areas with the same number of function sub-categories but different compositions. 

To compare different station areas, the indices were converted to the effective number (\cite{jost2006entropy}) or the equivalent number of equally rare species to the group of interest. Furthermore, the indices were expressed as a relationship of the percentage to the maximum diversity (26 for both indices in this study: equivalent to the number of sub-categories) to facilitate interpretation.

The Simpson's diversity-based index is defined as follows:
\begin{equation}
    B_{rl} = \frac{1}{\sum_{i} p_i^2} \times \frac{1}{\beta}_{max}
    \label{eq:b_rl}
\end{equation}
where ${\beta}_{max}$ equals the number of individuals $m$.

The Shannon's entropy-based index is defined as follows:
\begin{equation}
    H_{rl} = exp(\sum_i -p_i \ln{p_i}) \times \frac{1}{H}_{max}
    \label{eq:shannon}
\end{equation}
where ${H}_{max}$ also equals the number of individuals $m$. The value ranges from 0, which pertains to only one type of individual, to $m$, the number of individuals.

\textbf{Density}: Density is defined as the number of facilities located in the station area of a given radius, divided by the area of the station area. The authors use this definition to consider the \textit{status quo} of the facilities' location and the history of the areal development. It is possible to use the buildable area around a station as the denominator after considering the geographical constraints, such as rivers or coastlines; however, this operation implicitly assumes that the facilities would have been located over unbuildable areas if there were no obstacle. This assumption ignores the area's geographical characteristics and may lead to an overestimation of the facilities' density from visitors' perspectives. Therefore, the authors use the area of the circle with the given radius uniformly as the denominator of the density variable.

\subsection{Monte Carlo simulation}
The known properties of the diversity indices (e.g., their different sensitivity to sample size (\cite{Morishita1996})) have largely been studied using biological abundance distribution and may not apply to urban settings because the abundance distribution of facilities may differ from a biological one (e.g., log-normal distribution (\cite{lande1996statistics}) or the Zipf$-$Mandelbrot model (\cite{mouillot1999comparison})). To investigate the sample-size sensitivity of the diversity indices and to provide a baseline to compare with the observed correlation between diversity and density (explained in section \ref{sec:cat_proc}), the Monte Carlo simulation was conducted by producing a randomly sampled set of facilities in a station area.

% Outline of simulation
First, the authors constructed an empirical distribution of all facilities across the aggregated geographical area of all the station areas that were examined, reflecting the actual state of the facility distribution in an urban setting. One iteration involved sampling a set of facilities in a station area from a predefined population following the empirical distribution and calculating and recording the diversity indices. The total number of facility types was limited to 26 (discussed in detail in section \ref{sec:cat_def}). For each iteration, the number of facilities in the set began at 25 and increased in increments of 25 up to 1,000 after every thousandth iteration. The iterative sampling process for a number of facilities allowed for the estimation of the mean and 95\% confidence interval of the diversity indices of a hypothetical station area that contains the corresponding number of facilities. In addition, changing the sample sizes enabled observations of how the sample size influences diversity indices.

\subsection{Analytical procedure}
\label{sec:cat_proc}
The spatial data of the stations were projected to the local coordinate system (EPSG 2451) and the 400 m radius (\cite{cervero1997travel}) of the station areas was measured from the geometrical center of station platforms. Duplicate station names and operators in the original dataset were manually removed on the open-source geographic information system QGIS. Subsequently, facilities located inside each station area were extracted, and the diversity indices were calculated based on the number of facilities in the sub-categories. % database construction

The number and composition of facilities that are located near the stations are expected to be diverse because the stations' daily ridership varies from several hundreds to more than 1.5 million. To overcome these large differences in the scale of stations, the authors adopted the k-means clustering method. The station areas were clustered by the number of facilities and diversity indices, thereby enabling a comparison of station areas in each cluster. The average silhouette score was also employed to determine the number of clusters (\cite{rousseeuw1987silhouettes}); the cluster count with the highest score meaning the best separation of the individuals.

Finally, the correlation between the diversity and number of facilities (equivalent to density because all station areas have equal size) was estimated for each cluster. The objective of this process was to test the hypothesis that the diversity and density of facilities are associated in a group of station areas that evince certain similarities. Because these two variables deviated from normality (tested via the Shapiro-Wilk test; see the results in section \ref{section:result_cls}), Spearman's rank correlation coefficient was calculated and the value is denoted as $\rho$ (rho) hereinafter. %Furthermore, the result was compared to that of the Monte Carlo simulation to infer the structure of the facilities' locations. % Compare the result with that of the randomized simulation, and infer the structure of the facility location

\subsection{Robustness checks}
The robustness of the analytical findings was checked by varying the radius of station areas. From a geographic perspective, the authors also analyzed the station areas with different radius sizes while retaining other settings to verify the size sensitivity of the findings in the 400 m radius case. The alternative radius sizes were 600 m and 800 m, which corresponds to the approximate distance of a 7.5- and 10-minute walk (at 80 $m/min$; \cite{bivina2020walk}). The unique size of station areas by radius was imposed regardless of stations' scale because the robustness check was performed to estimate the general relation between the diversity and density of facilities, rather than to understand the differences in the geometrical expansion of agglomerated facilities in individual station areas. In addition, for these two cases, the same cluster counts as that for the 400 m radius was adopted to facilitate subsequent interpretations thereof; applying different cluster counts by radius makes the meanings of each cluster inconsistent from one radius to another.

%%%%%%%%%%%%%%%%%%%%%%%%%%%%%%%%%%%%%%%%%%
\section{Results}
\label{section:4}
%%%%% 12 + 12 paragraphs (Computer, Environment and Urban Systems)%%%%%
\subsection{Data description}
\begin{figure}[h]
    \centering
    \includegraphics[scale=0.5]{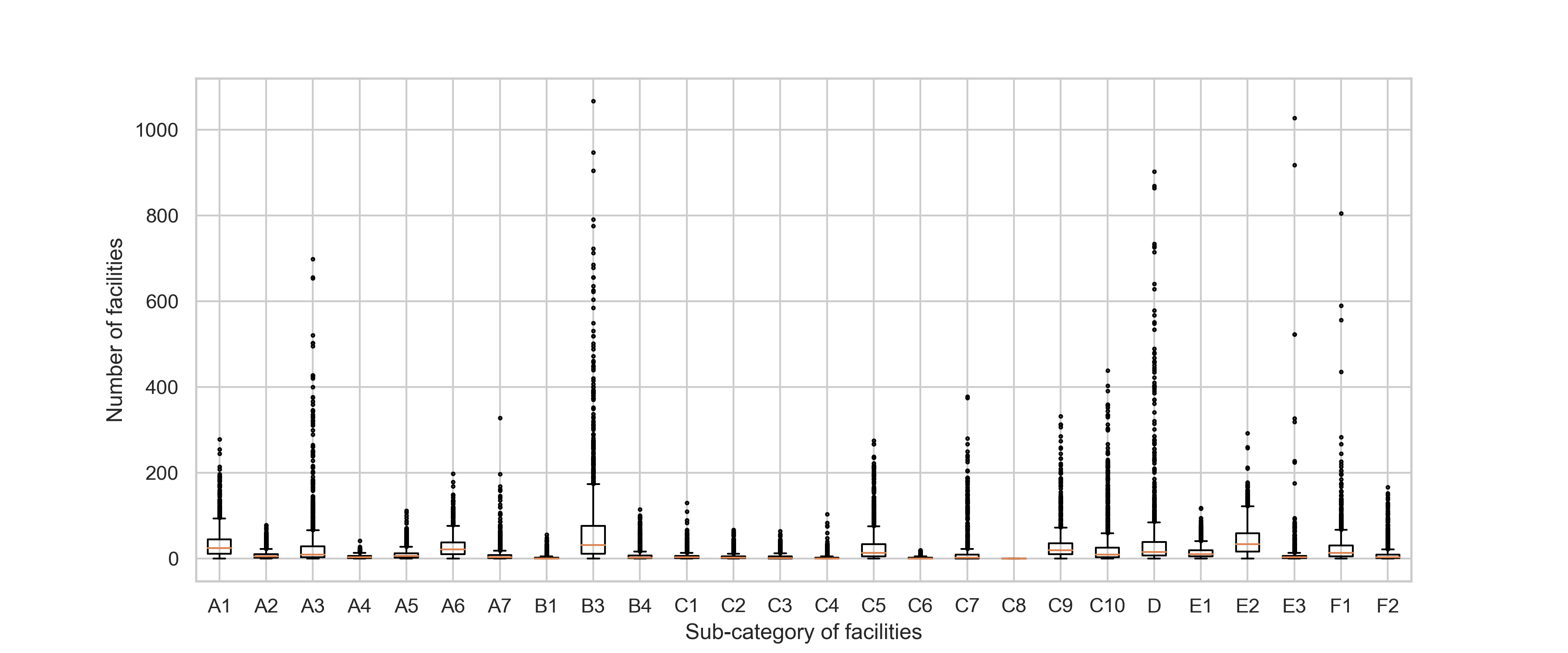}
    \caption{Boxplot of the number of facilities by the facility sub-categories. Each point represents a station area (400 m radius).}
    \label{apd_img:stats_boxplot}
\end{figure}

Figure \ref{apd_img:stats_boxplot} illustrates the number of each sub-category in a boxplot, revealing that some sub-categories are more abundant than others (e.g., A1, A6, B3, C9, and E2). The variance of some sub-categories was also larger than those of others, implying a non-normal distribution in the number of facilities in the sub-categories.

\subsection{Numerical behavior of diversity indices}
\begin{figure}[h]
    \centering
    \includegraphics[scale=0.35]{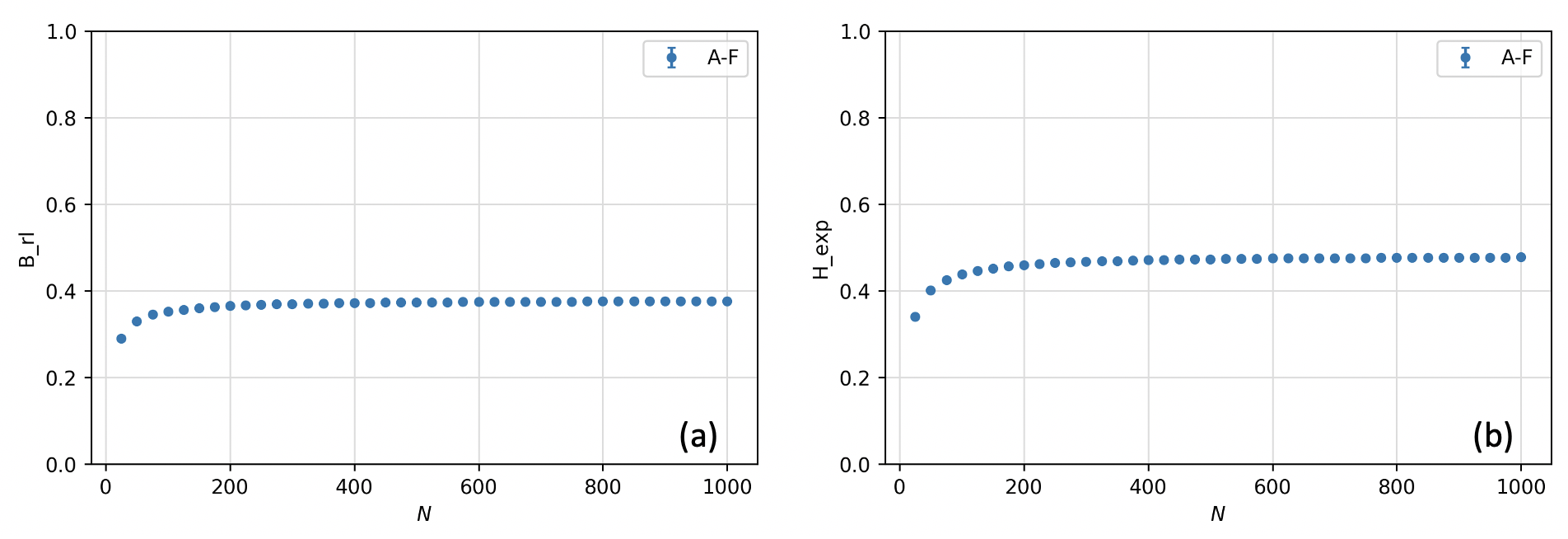}
    \caption{Sample-size sensitivity of (a) Simpson's diversity-based index and (b) Shannon's entropy-based index. The sample was sourced from an empirical distribution of facilities across the entire study area.}
    \label{img:simu}
\end{figure}

Figure \ref{img:simu} illustrates the sample-size sensitivity of the diversity indices from the Monte Carlo simulation. This showed that, first, both indices increased their values monotonically as density increased, and second, the values were stable where the density was not low (e.g., N$\geq$200). The first point suggests that the indices would increase if the facilities were randomly distributed over station areas. This observation serves as a baseline in comparison to the empirical analysis that follows. The second point, in turn, justifies the comparison between station areas with different densities because the measured diversity would have contained bias from the corresponding density if the diversity indices were sensitive to the density change. These points validate the usage of these indices in arguing the relation between diversity and density of urban functions.

% The first point suggests that the negative correlation between the relative diversity and density observed in the high cluster was not derived from the measurement of the diversity but from the spatial distribution of the facilities.
% These points, hence, corroborates the finding in the study area that highly densified station areas are less diverse in terms of facility types, measured with the relative diversity.

% \clearpage
\subsection{Station areas' characteristics by cluster}
\label{section:result_cls}
\begin{figure}[h]
    \centering
    \includegraphics[scale=0.35]{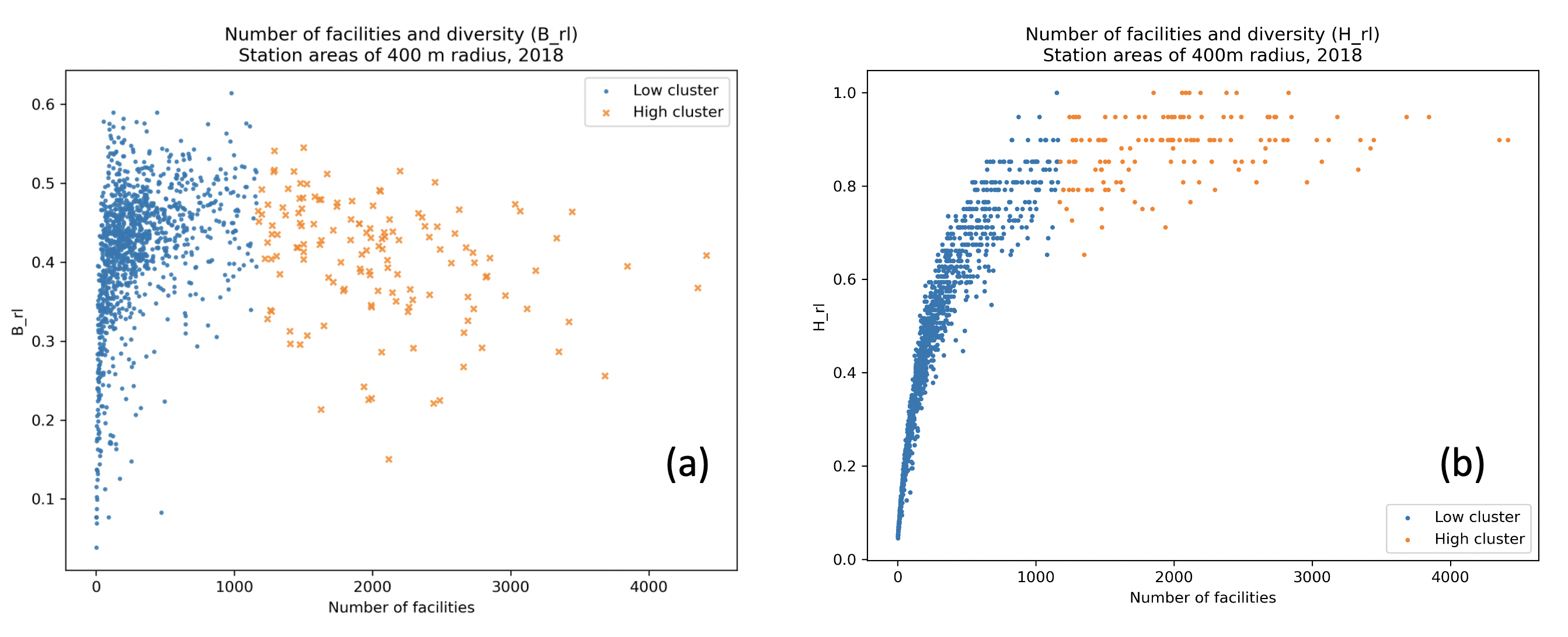}
    \caption{Results of the clustering. Number of facilities and diversity (all sub-categories) of station areas (400 m radius) measured by (a) Simpson's diversity-based index and (b) Shannon's entropy-based index.}
    \label{img:cluster3_B_rl_400}
\end{figure}

Figure \ref{img:cluster3_B_rl_400} shows the scatterplot of the number and diversity of the facilities of all sub-categories that were located within a 400 m radius of the stations studied, measured by Simpson's diversity-based index and Shannon's entropy-based index. The cluster counts were determined to 2 because the average silhouette scores were higher (0.79) than other counts (Figure S1 in the supplementary material).

\begin{table}[h]
\centering
\begin{tabular}{r|cc|cc}
                                      & \multicolumn{2}{c}{Simpson's diversity (B\_rl)} & \multicolumn{2}{c}{Shannon’s diversity (H\_rl)} \\ \hline
Cluster                               & Low       & High            & Low       & High            \\ \hline
Number of facilities & $\leq$ 1,160             & \textgreater{}1,160       & $\leq$ 1,160             & \textgreater{}1,160        \\
Mean                 & 298                & 2034                      & 298                & 2034                       \\ \hline
Diversity value      & 0.038-0.614        & 0.150-0.546               & 0.045-1.00        & 0.653-1.00                \\
Mean                 & 0.413              & 0.404                     & 0.468              & 0.879                      \\ \hline
Spearman’s rho       & 0.479              & -0.328                    & 0.973              & 0.375                      \\
p-value              & 0.000              & 0.004                     & 0.000              & 0.000                      \\ \hline
N                    & 1,212              & 140                       & 1,212              & 140                        
\end{tabular}
\caption{Characteristics of each cluster with cluster count 2 (400 m radius).}
\label{tab:cls2_BH}
\end{table}

% different trend according to cluster
To test the hypothesis on the association between the diversity and density of facilities, correlations in each cluster were examined. Following a Shapiro-Wilk test, the null-hypothesis (i.e., the samples are drawn from a normal distribution) was rejected for all three cases with a significance level of 5\%. This necessitated the calculation of Spearman's rank correlation; Table \ref{tab:cls2_BH} shows the results thereof.

For both indices, the low clusters (station areas with low facility density) had wider ranges of diversity value and showed positive correlations between diversity and density—$\rho$ = 0.479 and 0.973, respectively—which were also observed in the Monte Carlo simulation. However, regarding Simpson's diversity-based index, the simulation did not show the negative coefficient ($\rho$ = $-$0.328) of the high cluster (station areas with high facility density). If the facilities were randomly located in all station areas, the coefficient would have been at least slightly positive. This result indicates that a high density of facilities was associated with a loss in terms of diversity, which is a reversal of the initial hypothesis. Shannon's entropy-based index, in turn, was still positively correlated with the density of facilities in the high cluster. An interpretation for this difference is discussed in section \ref{section:5}.

\begin{figure}[h]
    \centering
    \includegraphics[scale=0.5]{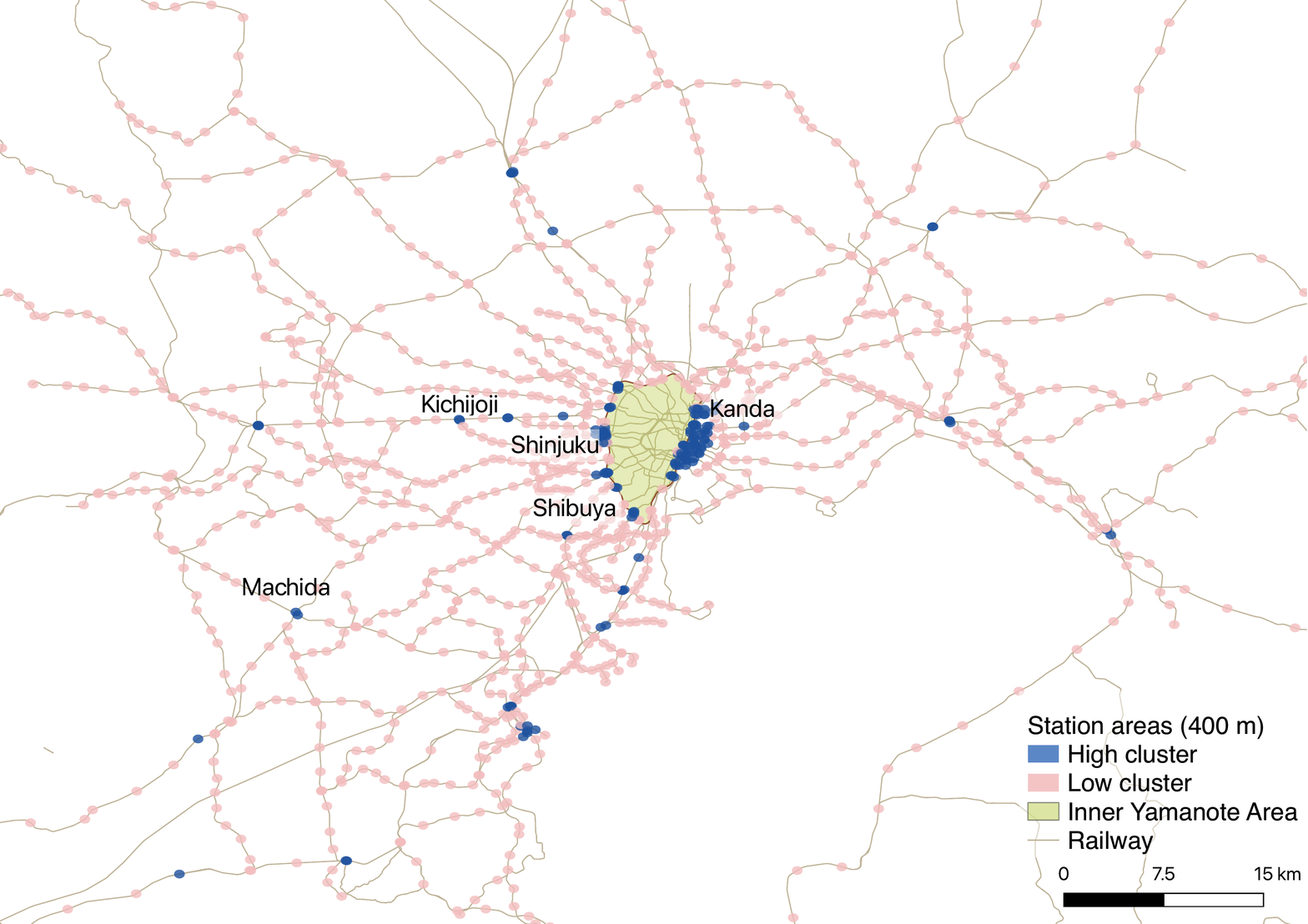}
    \caption{Station areas in the Tokyo Metropolitan Area (400 m radius). Clustering based on Simpson's diversity-based index and density of facilities.}
    \label{img:station_areas_all}
\end{figure}

The station areas are plotted on a map in Figure \ref{img:station_areas_all}, with the clustering based on Simpson's diversity-based index. Highly dense station areas (high cluster) were concentrated on the termini along the Yamanote loop line (at the fringe of the green area in the figure) as well as other hub stations in the suburban area, which typically have railway connections. Conversely, lowly dense station areas (low cluster) were primarily located around stations with few railway connections.

\clearpage
\subsection{Robustness checks}
\begin{figure}[h]
    \centering
    \includegraphics[scale=0.4]{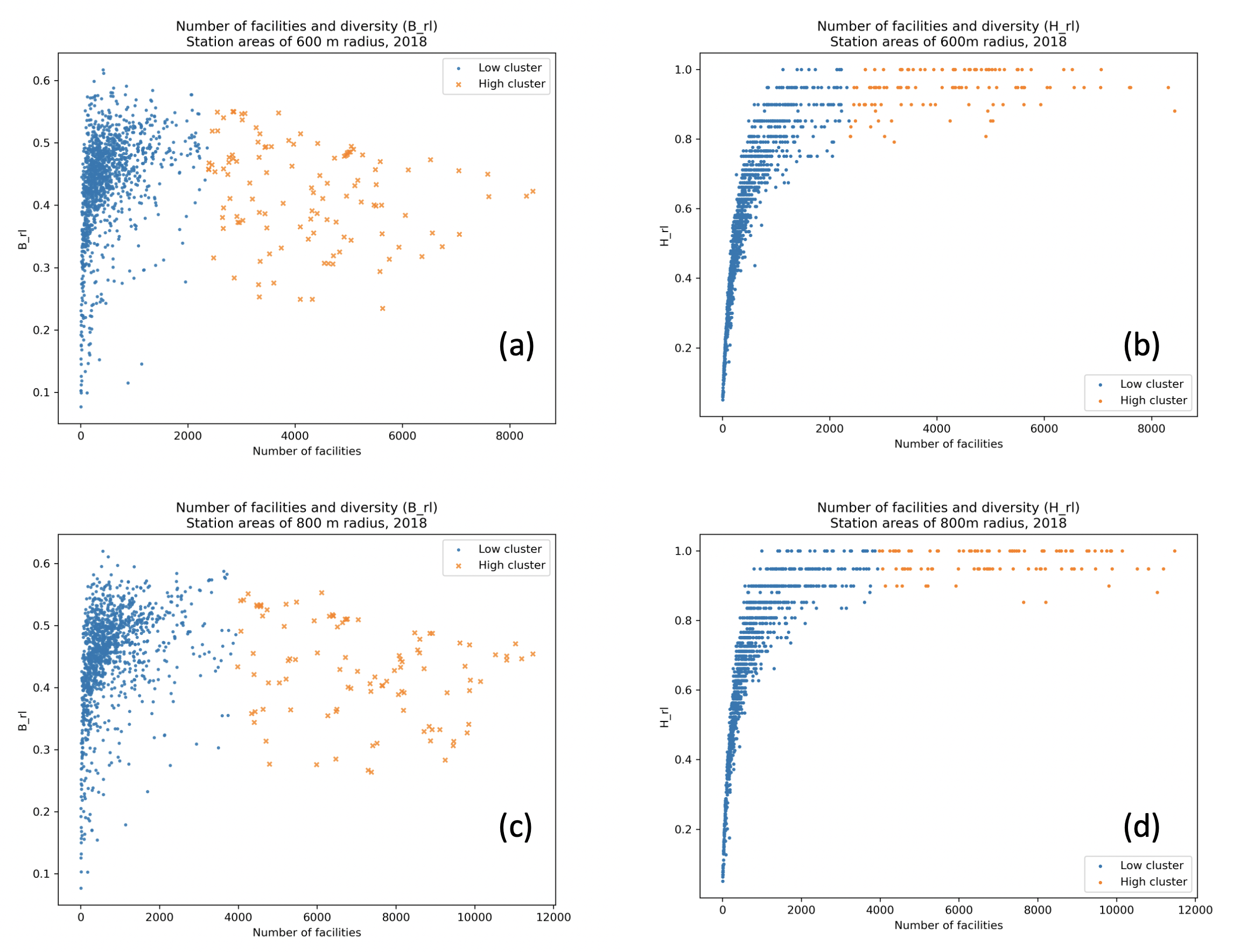}
    \caption{Results of the sensitivity analysis. Number of facilities and diversity (all sub-categories) of station areas (Simpson's diversity-based index for (a) 600 m and (c) 800 m radius and Shannon's entropy-based index for (b) 600 m and (d) 800 m radius).}
    \label{img:cluster3_B_rl_600800}
\end{figure}

The reported results were limited to a 400 m radius area around stations. To discuss whether the findings are generalizable at a wider scale, analyses on different radii were conducted. Figure \ref{img:cluster3_B_rl_600800} presents the scatterplots of the individual station areas, and Table \ref{tab:spearman_800} shows Spearman's $\rho$ for the station areas in each cluster. As with the 400 m radius case for Simpson's diversity-based index, significant positive correlations in the lowly dense station areas and negative positive correlations in the highly dense station areas were observed in both the 600 m and 800 m radii cases. The degree of the negative correlation varied depending on the radius. Conversely, in the case of Shannon's entropy-based index, only the highly dense station areas behaved differently from the others, showing no significant correlation between the diversity of facilities and their density.

Some station areas belonged to different clusters depending on the radius. For example, Machida was in the highly agglomerated cluster for the 400 m radius but in the lowly agglomerated cluster for the 600 m and 800 m radii. This may be because of the different degrees of spatial expansion of the facilities' locations between station areas. Nonetheless, the results indicate that a similar tendency of association between the diversity and density of facilities is present regardless of geographical scale.

\begin{table}[h]
\centering
\begin{tabular}{cr|cc|cc}
\multicolumn{1}{l}{}       & \multicolumn{1}{l}{} & \multicolumn{2}{c}{Simpson's diversity (B\_rl)} & \multicolumn{2}{c}{Shannon’s diversity (H\_rl)} \\ \hline
\multicolumn{1}{l}{Radius} & Cluster              & Low                    & High                  & Low                           & High                          \\ \hline
\multirow{2}{*}{600m}      & Spearman’s rho       & 0.486                  & -0.263                & 0.955                         & 0.282                         \\
                           & (p-value)            & (0.000)                & (0.004)               & (0.000)                       & (0.002)                       \\ 
\multirow{2}{*}{800m}      & Spearman’s rho       & 0.487                  & -0.292                & 0.945                         & 0.066                         \\
                           & (p-value)            & (0.000)                & (0.003)               & (0.000)                       & (0.509)                      
\end{tabular}
\caption{Spearman's rank correlation (rho) in each cluster (600 m and 800 m radius, Simpson's diversity-based index and Shannon's entropy-based index).}
\label{tab:spearman_800}
\end{table}

%%%%%%%%%%%%%%%%%%%%%%%%%%%%%%%%%%%%%%%%%%
\section{Discussion}
\label{section:5}
\subsection{Results based on diversity indices}
A high density of facilities was associated with low diversity if the relation was measured with Simpson's diversity-based index, but not with Shannon's entropy-based index. A reason for this difference comes from the control parameter \textit{q} in the Hill number: the Simpson's diversity-based index ($q=2$) weights the dominant species more than Shannon's entropy-based index ($q=1$) does. Mathematically speaking, Shannon's entropy-based index has an additive property (\cite{keylock2005simpson}) that an increase in the number of considered individuals necessarily increases the index value; this can be interpreted as the equal weighting of species. Conversely, Simpson's diversity-based index has a sub-additive property that discounts the impact of an increase in the number of considered individuals on the index value; this corresponds to a weighting for dominant species.
% While both indices satisfy the desirable conditions for a diversity index (XX), the relative diversity is more appropriate to the purpose of this study than the other for the following reason.

Considering the scope of this study, Simpson's diversity-based index is more appropriate because it represents the diversity of facilities evaluated by visitors in a station area. It is probable that they evaluate the degree of diversity in the area based on dominant species (functions) that they likely see and may not consider all the functions present in the area as equal. In other words, the Shannon's entropy-based index overestimates functions that are barely recognized by visitors. Additionally, it has the same degree of weighting for rare species as the Hirschmann-Herfindahl index (\cite{yue2017measurements}). As the HHI is used to measure industrial the supply-side diversity on an entity basis, it is more consistent with previous works to use Simpson's diversity-based index.

It should be noted that the declining phase of the relationship between diversity and density measured with Simpson's diversity-based index is not caused by the innate property of the index, as demonstrated by the Monte Carlo simulation. Therefore, the negative correlation can be considered as a result of the non-random location of the functions, such that the diversity of functions tends to be low where density is high.

\subsection{Mechanism guiding the association between diversity and density}
A possible explanation for the negative correlation relates to the non-random spatial distribution of the facilities as demonstrated by the difference between the empirical results and the Monte Carlo simulation results. Along with the number of facilities in the station areas, some station areas tended to accommodate certain types of facilities in terms of the sub-categories, whereas others did not. This difference may derive from location strategies of individual facilities: for instance, the strategy of a retail store around a railway station is likely different from that of a factory in the same area owing to land price, land-use regulations, and other factors (\cite{fujita1996economics}). The relatively high cost of housing near a station discourages certain types of facilities that require a broad floor area or cannot afford such costs, thus partly limiting the variety of facilities in the station area. Furthermore, the products and services that the facilities sell and offer determine their geographical catchment area (e.g., order of products; \cite{christaller1966central}), which also contributes to the heterogeneous distribution of facilities in a collective scale. The catchment areas of grocery stores (A1), for instance, are more likely to be narrower than those of leisure-related facilities (A5) (e.g., movie theaters) and much more so than those of educational facilities (E3) (e.g., high schools). Because facilities with wide catchment areas tend to be located more sparsely than those with narrow ones, a geographically delimited area will have more of the former facilities than the latter.

Another possible explanation relates to a lack of necessity for diversification of urban functions because of high accessibility to other areas. In other words, a region composed of station areas with high mutual accessibility may be able to satisfy people's demand at the region level and not at the individual station area level. If a geographical area offers no motorized or public transport modes, it would develop to contain a sufficient variety of functions (facilities), thereby improving its diversity value. Conversely, if an area is accessible from other areas via transport connections, it may be sufficient for the region, composed of several areas, to have a variety of facilities, without each area in the region having diversified urban functions in itself, because residents can easily meet their demand by visiting neighboring areas. The representation in Figure \ref{img:station_areas_all} supports this view, showing that station areas of the high cluster are accessible from multiple railway lines, typically concentrated on the fringes of the inner Yamanote area, which is the core of the Tokyo Metropolitan Area. The high accessibility among station areas may be associated with high density and lower diversity of urban functions than those with moderate density.  % This observation suggests that the accessibility between station areas contributes to the relation between diversity and density of urban facilities.

A comparison between Kanda (B\_rl = 0.324, N = 3,420) and Kichijoji (B\_rl = 0.446, N = 2,377), both classified as highly dense station areas within the 400 m radius, supports this view. Kichijoji has relatively abundant facilities in category A (retail commerce (A1$-$A7), 40.9\% in total), whereas the corresponding proportion in Kanda is low (9.9\%). In addition, D (gross commerce factories) is the most dominant in Kanda (25.4\%), followed by B3 (restaurants, 13.5\%), whereas A3 (stores for personal equipment, 14.1\%) is as dominant as B3 (16.8\%) in Kichijoji. The relatively low supply of category A in Kanda would be compensated by other nearby station areas such as Akihabara (B\_rl = 0.402, N = 2,106; ratio of category A is 20.0\%) and Ueno (B\_rl = 0.441, N = 1,716; category A: 27.0\%), and Kanda would reciprocally provide a supply of category D. By contrast, Kichijoji, which has less accessibility to other station areas, provides diversified functions on its own including those in category E (services of education (E1), health and social assistance (E2), and public administration (E3); 10.6\% in total); the corresponding proportion in Kanda is only 5.1\%.

\subsection{Policy implications}
Station areas with low diversity in terms of their functions are more vulnerable to external shocks such as economic demand (e.g., recessions, natural disasters, and pandemics such as COVID-19), as indicated in portfolio theory (cf. \cite{montgomery1994corporate,frenken2007related}). Portfolio theory states that a firm that sells products with uncorrelated demand reduces the risk of a sudden decrease in a certain type of demand more effectively than a firm that sells products with correlated demand because the loss is compensated by profits from other products. The same holds for urban functions focusing on economic sectors; namely, a station area with diverse facilities may remain functional and attractive to citizens despite a decrease in a certain category's economic demand (cf. \cite{wink2012economic}). Therefore, high diversity of urban functions is desirable for both urban vibrancy and resilience to demand shocks.

% exclusionary facilityの例を入れる
Given the importance of diversity, the findings of this study suggest that highly dense station areas require development strategies that maintain or improve diversity in the areas because diversity and density of urban functions are negatively correlated. An example of such strategies is the financial support to the facilities of an under-represented category in an area, which could help diversify the functions in the area. Another example is imposing restrictions on exclusionary facilities to be located in an area. Because some types of facilities (i.e., large-scale shopping centers) are exclusionary, preventing other facility types from being located nearby or controlling the number of such facilities may facilitate the opening and maintenance of other facilities that would have otherwise been excluded. Nonetheless, the legitimacy of these interventions to the land market must be further studied from theoretical and empirical perspectives.

Regarding the station areas with a small or moderate number of facilities, however, urban planners may not have to prioritize diversity over density as much as they would for station areas with numerous facilities. This is because the empirical results in this study show that the diversity and density, both being sources of urban vibrancy, have a positive association. Therefore, it is likely sufficient to attract firms or facilities to the areas without limiting the categories, such as by investing in the infrastructure for example. Nevertheless, further understanding of the causality between the two is of vital importance for designing an effective development strategy.

\subsection{Limitations}
This study has three limitations. First, it ignored the heterogeneity between facilities other than their functional sub-categories; for instance, a facility with a large floor area and another with a small one were treated equally in the data set. The bias coming from this limitation could be attenuated by using a more detailed data set about the facilities. Second, the analysis herein only considered non-residential aspects of cities; consequently, it could not account for the location patterns of residents and their magnitudes. These aspects can influence the facilities' locations. Considering this demand-side data allows to produce more insights for location theories, such as the optimal mixture of facilities under a given demand of residents in an area. Third, it did not examine the causality between the diversity and density of urban functions, because of the absence of a longitudinal dataset. Given that each geographical area began from a small assemblage of facilities, such a dataset would quantify the dynamic aspect of the patterns of the facilities' location, for example, whether a higher density precedes a lower diversity at highly dense station areas.

%%%%%%%%%%%%%%%%%%%%%%%%%%%%%%%%%%%%%%%%%%
\section{Conclusion}
\label{section:6}
The diversity of urban functions around station areas was found to be associated with their density. In particular, lower density areas tended to show a positive correlation between the density and diversity of facilities, whereas higher density ones tended to show a negative correlation. This tendency was observed in all the radii examined, implying a generality of the association.

The results of this paper and the importance of the diversity of functions should encourage policy-makers, urban planners, and developers to stimulate the diversity—and not merely the density—of urban functions, particularly in highly dense station areas. Diversifying urban functions in a given area would require an intentional effort because large-scale development plans often include large buildings that accommodate a limited variety of functions. If built upon quantitative observations, such efforts could result in a vibrant and resilient area.

%%%%%%%%%%%%%%%%%%%%%%%%%%%%%%%%%%%%%%%%%%
% \vspace{6pt} 

%%%%%%%%%%%%%%%%%%%%%%%%%%%%%%%%%%%%%%%%%%
%% optional
%\supplementary{The following are available online at \linksupplementary{s1}, Figure S1: title, Table S1: title, Video S1: title.}


%%%%%%%%%%%%%%%%%%%%%%%%%%%%%%%%%%%%%%%%%%
% \authorcontributions{conceptualization, Y.K. and Y.Y.; methodology, Y.K., S.C and Y.Y.; software, Y.K., S.C. and X.L.; validation, Y.K.; formal analysis, Y.K.; investigation, Y.K.; resources, Y.K., S.C. and X.L.; data curation, Y.K. and S.C.; writing--original draft preparation, Y.K.; writing--review and editing, S.C., H.K., X.L. and Y.Y.; visualization, Y.K.; supervision, Y.Y.; project administration, Y.Y.; funding acquisition, Y.Y.. All authors have read and agreed to the published version of the manuscript.}

%%%%%%%%%%%%%%%%%%%%%%%%%%%%%%%%%%%%%%%%%%
% \acknowledgments{test}
\section*{Acknowledgments}
The authors are grateful to two anonymous reviewers for their comments and to Mr. Kantaro Yamaguchi, Mr. Otoya Kobayashi, Mr. Yosuke Isobe, Dr. Fumihiko Omori and Mr. Atsushi Omachi from Tokyu Corporation for insightful discussions. This research was the result of a joint research with CSIS, the University of Tokyo (No. 986) and used the following data: Telepoint Pack DB (Yellow Pages) provided by Zenrin CO..

% %%%%%%%%%%%%%%%%%%%%%%%%%%%%%%%%%%%%%%%%%%
% \section*{Declaration of conflicting interests}
% The author(s) declared no potential conflicts of interest with respect to the research, authorship and/or publication of this article.
% % \conflictsofinterest{The authors declare no conflict of interest.}
% % \conflictsofinterest{Declare conflicts of interest or state ``The authors declare no conflict of interest.'' Authors must identify and declare any personal circumstances or interest that may be perceived as inappropriately influencing the representation or interpretation of reported research results. Any role of the funders in the design of the study; in the collection, analyses or interpretation of data; in the writing of the manuscript, or in the decision to publish the results must be declared in this section. If there is no role, please state ``The funders had no role in the design of the study; in the collection, analyses, or interpretation of data; in the writing of the manuscript, or in the decision to publish the results''.} 

%%%%%%%%%%%%%%%%%%%%%%%%%%%%%%%%%%%%%%%%%%
% \funding{This research received no external funding}
% \funding{}
% \section*{Funding}
% This research received no external funding.

% %%%%%%%%%%%%%%%%%%%%%%%%%%%%%%%%%%%%%%%%%%
% %% optional
% \appendixtitles{yes} %Leave argument "no" if all appendix headings stay EMPTY (then no dot is printed after "Appendix A"). If the appendix sections contain a heading then change the argument to "yes".
\clearpage
\appendix
\section{Categories of facilities}

\begin{table}[h]
\begin{tabular}{lll}
\textbf{Category}                                             & \textbf{Sub-category}               & \textbf{Activity}                                                                    \\ \hline
A (Retail commerce)                                  & A1 & Groceries                                                                   \\
                                                     & A2                         & Housing equipment                                                      \\
                                                     & A3                         & Personal equipment                            \\
                                                     & A4                         & Automobile                                                                  \\
                                                     & A5                         & Hobby and leisure                                                           \\
                                                     & A6                         & Other services                                                              \\
                                                     & A7                         & Shopping center, department store                                           \\ \hline
B (Local non-tradable service)                       & B1                         & Hotel, B\&B, hostel                                                         \\
                                                     & B3                         & Restaurant                                                                  \\
                                                     & B4                         & Bar, caf\'{e}                                                               \\ \hline
C (Tertiary sector of industry)                      & C1                         & Logistics and transport                                                     \\
                                                     & C2                         & Bank, finance                                                               \\
                                                     & C3                         & Insurance                                                                   \\
                                                     & C4                         & Financial intermediation                                                    \\
                                                     & C5                         & Real estate                                                                 \\
                                                     & C6                         & Telecommunication                                                           \\
                                                     & C7                         & IT activities                                                               \\
                                                     & C8                         & R\&D                                                                        \\
                                                     & C9                         & 
                                        \begin{tabular}[l]{@{}l@{}}Architecture, civil engineering, publication\\and other specialized services\end{tabular}
                                                      \\
                                                     & C10                        & Other services to people                                                    \\ \hline
D (Gross commerce)                                   & D                          & Gross commerce                                                              \\ \hline
E (Public services)                                  & E1                         & Services of education                                                       \\
                                                     & E2                         & Health and social assistance                                                \\
                                                     & E3                         & Public administration                                                       \\ \hline
F (Association, recreational and sport activities) & F1                         & Recreational and cultural and sport activities                              \\
                                                     & F2                         & Association activities                                                   
\end{tabular}
\caption{Categories and the corresponding activities}
\label{appendix:categories}
\end{table}

\end{document}